\documentclass[fleqn,twoside]{article}
\usepackage{espcrc2}

\usepackage{epsfig}

\usepackage{psfrag}
\newcommand{\cha}{\tilde{\chi}}
\newcommand{\neu}{\tilde{\chi}^0}
\def\SPHENO{SPheno\,2.2.0}

\title{SUSY Parameter Analysis at TeV and Planck Scales}

\author{
B.~C.~Allanach\address{LAPTH, Annecy-le-Vieux, France}, 
G.~A.~Blair\address[DESY]{Deutsches Elektronen-Synchrotron DESY,
                          D-22603 Hamburg, Germany}
           \address{Royal Holloway University of London, Egham, Surrey. TW20 0EX,
                    UK}, 
A.~Freitas\address{Fermi National Accelerator Laboratory, P. O. Box 500, 
                                                     Batavia IL 60510, USA},
S.~Kraml\address{Inst. f. Hochenergiephysik, \"Osterr. Akademie d. 
                  Wissenschaften, A-1050 Vienna, Austria}
        \address[CERN]{CERN, Department of Physics, CH-1211 Geneva 23, 
                                                                Switzerland}, 
H.-U.~Martyn\address{I. Physik. Institut, RWTH Aachen, D-52074 Aachen, Germany},
G.~Polesello\addressmark[CERN],
W.~Porod\address{IFIC - Instituto de F\'\i sica Corpuscular,
                 E-46071 Valencia, Spain},
and P.~M.~Zerwas\addressmark[DESY]
} 

\begin{document}

\begin{abstract}
Coherent analyses at future LHC and LC
experiments can be used to explore the breaking mechanism of supersymmetry
and to reconstruct the fundamental theory at high energies, in particular
at the grand unification scale. This will be exemplified for minimal
supergravity.
\end{abstract}

\maketitle

\section{Physics Base}

The roots of standard particle physics are expected to go as 
deep as the Planck length of $10^{-33}$~cm where gravity is 
intimately linked to the particle system.  A stable bridge
between the electroweak energy scale of 100~GeV and
the vastly different Planck scale of $\Lambda_{\rm PL}\sim10^{19}$~GeV,
and the (nearby) grand unification scale $\Lambda_{\rm GUT}\sim10^{16}$~GeV,
is provided by supersymmetry. Methods must therefore be developed which allow to
 study the 
supersymmetry breaking mechanism and the 
physics scenario near the GUT/PL scale \cite{r1}.

The reconstruction of physical structures at energies more than
fourteen orders above accelerator energies is a
demanding task. LHC~\cite{r2} and a future e$^+$e$^-$ linear collider
(LC)~\cite{r3} are a perfect tandem for solving such a problem:
 While the colored supersymmetric particles, gluinos and squarks,
can be generated with large rates for masses up to 2 to 3~TeV at the
LHC, the strength of e$^+$e$^-$ linear colliders is the comprehensive
coverage of the non-colored particles, charginos/neutralinos and
sleptons.  If the analyses are performed coherently, the accuracies in
measurements of cascade decays at LHC and in threshold production as
well as decays of supersymmetric particles at LC complement each other
mutually. A comprehensive and precise picture is needed
in order to carry out the evolution of the supersymmetric parameters
to high scales, which is driven by perturbative loop effects involving 
the entire supersymmetric particle spectrum.

Minimal supergravity [mSUGRA] provides us with a scenario within
which these general ideas can be quantified. 
Supersymmetry is broken in mSUGRA in a hidden sector and the breaking
is transmitted to our eigenworld by gravity~\cite{Chamseddine:jx}.  
The mechanism suggests, yet does not enforce [see {\it e.g.} 
Ref.\cite{Wells}], the universality of the
soft SUSY breaking parameters -- gaugino and scalar masses, 
trilinear couplings -- at a scale that is generally identified with
the unification scale. Alternative scenarios have been formulated
for left--right symmetric extensions, superstring effective theories,
and for other SUSY breaking mechanisms.

\section{Minimal Supergravity}

\begin{figure}[t]
\begin{center}
\setlength{\unitlength}{1mm}
\begin{picture}(75,50)
\psfrag{h}{\Huge $h^0$}
\psfrag{J}{\Huge $H^0$}
\psfrag{Z}{\Huge $\,A^0$}
\psfrag{I}{\Huge $H^\pm$}
\psfrag{N}{\Huge $\neu_1$}
\psfrag{M}{\Huge $\neu_2$}
\psfrag{O}{\Huge $\cha^\pm_1$}
\psfrag{P}{\Huge $\tilde g$}
\psfrag{Y}{\Huge $\neu_3$}
\psfrag{X}{\Huge $\tilde{\tau}_1$}
\psfrag{T}{\Huge $\tilde{\tau}_2$}
\psfrag{R}{\Huge $\tilde{e}_{\rm R}$}
\psfrag{L}{\Huge $\tilde{e}_{\rm L}$}
\psfrag{V}{\Huge $\tilde{\nu}_\tau$}
\psfrag{U}{\Huge $\tilde{\nu}_e$}
\psfrag{B}{\Huge $\tilde{t}_1$}
\psfrag{A}{\Huge $\tilde{b}_1$}
\psfrag{C}[r][r]{\Huge $\tilde{q}_{\rm R}, \tilde{b}_2$}
\psfrag{D}{\Huge $\tilde{q}_{\rm L}$}
\psfrag{F}[r][r]{\Huge $\tilde{t}_2$}
\psfrag{Q}[r][r]{\Huge $\neu_4, \cha^\pm_2\!\!\!\!\!\!$}
\put(0,-15){%
\epsfig{figure=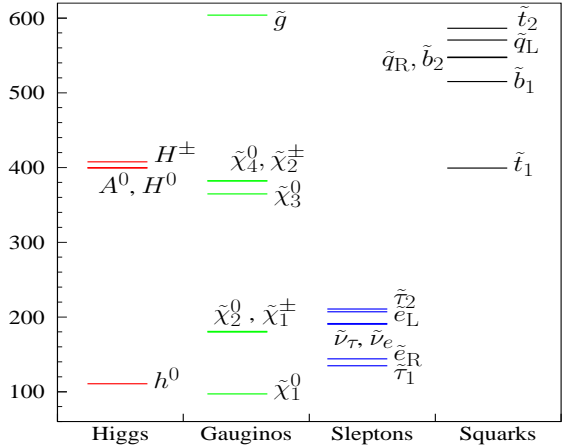, angle=0, width=7.5cm, height=6.2cm,
        viewport=32 0 540 505, clip=true}
}
\end{picture}
\end{center}
\caption{\it Spectrum of Higgs, gaugino/higgsino and sparticle masses
    in the mSUGRA scenario SPS1a [masses in {\normalfont GeV}].}
\label{fig:SPS1a_spectrum}
\end{figure}

The mSUGRA Snowmass reference point SPS1a is characterised by
the following values~\cite{r4} 
\begin{eqnarray}
  \begin{array}{ll}
    M_{1/2} = 250~{\rm GeV} \qquad & M_0=100~{\rm GeV} \\
    A_0=-100~{\rm GeV} & {\rm sign}(\mu)=+\\
    \tan\beta=10 & \\
  \end{array}
\label{eq:sps1a}
\end{eqnarray}
for the universal gaugino mass $M_{1/2}$, 
the scalar mass $M_0$, the trilinear coupling
$A_0$, the sign of the higgsino parameter $\mu$, and $\tan\beta$,
the ratio of the vacuum-expectation values of the two Higgs fields.
As the modulus of the higgsino parameter is fixed at the
electroweak scale by requiring
radiative electroweak symmetry breaking, $\mu$ is finally given by
$\mu = 357.4~{\rm GeV}$. 
The form of the supersymmetric mass spectrum of SPS1a is shown
in Fig.~\ref{fig:SPS1a_spectrum}.  In this scenario the squarks and gluinos
can be studied very well at the LHC while the non-colored gauginos
and sleptons can be analyzed partly at LHC and in comprehensive form
at an e$^+$e$^-$ linear collider operating at a total energy up to
1 TeV with high integrated luminosity close to 1~ab$^{-1}$.

At {\underline{LHC}} the masses can best be obtained by analyzing 
edge effects in the
cascade decay spectra. The basic starting point is the identification
of a sequence of two-body decays:
\mbox{$\tilde q_L\rightarrow\tilde\chi^0_2 q\rightarrow\tilde\ell_R\ell q
\rightarrow \tilde\chi^0_1\ell\ell q$}.
One can then measure the kinematic edges 
of the invariant mass distributions among the two leptons and the jet resulting
from the above chain, and 
thus an approximately model-independent determination 
of the masses of the involved sparticles is obtained \cite{HinPai,Cambr}.
The four sparticle masses [$\tilde q_L$, $\tilde\chi^0_2$,
$\tilde\ell_R$ and $\tilde\chi^0_1$] are used subsequently as input for
additional decay chains like
\mbox{$\tilde g\rightarrow\tilde b_1 b\rightarrow \tilde\chi^0_2 bb$},
and the shorter chains \mbox{$\tilde q_R\rightarrow q \tilde\chi^0_1$}
and \mbox{$\tilde\chi^0_4\rightarrow\tilde\ell\ell$}, 
which all require the 
knowledge of the sparticle masses downstream of the cascades.

\renewcommand{\arraystretch}{1.07}
\begin{table}
\begin{center}
\begin{tabular}{|c||c||c|c||c|}
\hline
   & Mass & ``LHC''  &\ ``LC''\ & ``LHC+LC''
\\ \hline\hline
$\tilde{\chi}^\pm_1$ & 179.7 &          & 0.55   &  0.55  \\
$\tilde{\chi}^\pm_2$ & 382.3 &     --   & 3.0    &  3.0   \\
$\tilde{\chi}^0_1$   &  97.2 &     4.8  & 0.05   &  0.05  \\
$\tilde{\chi}^0_2$   & 180.7 &     4.7  & 1.2    &  0.08  \\
$\tilde{\chi}^0_3$   & 364.7 &          & 3-5    &  3-5   \\
$\tilde{\chi}^0_4$   & 381.9 &     5.1  & 3-5    &  2.23  \\
\hline
$\tilde{e}_R$        & 143.9 &     4.8  & 0.05   &  0.05  \\
$\tilde{e}_L$        & 207.1 &     5.0  & 0.2    &  0.2   \\
\hline
$\tilde{q}_R$        & 547.6 &    7-12  &    --  & 5-11  \\
$\tilde{q}_L$        & 570.6 &     8.7  &    --  &  4.9  \\
$\tilde{t}_1$        & 399.5 &          & 2.0    &  2.0  \\
\hline
$\tilde{g}$          & 604.0 &     8.0  &    --  &  6.5  \\
\hline
$h^0$                & 110.8 &     0.25 & 0.05   & 0.05   \\
$H^0$                & 399.8 &          & 1.5    & 1.5   \\
$A^0$                & 399.4 &          & 1.5    & 1.5   \\
$H^{\pm}$            & 407.7 &     --   & 1.5    & 1.5   \\\hline 
\end{tabular}\\
\end{center}
\caption{{\it Accuracies for representative mass measurements
at ``LHC'' and ``LC'', cf. Ref.~\cite{LC-errors}, 
and in coherent ``LHC+LC'' analyses
for the reference point SPS1a [masses in {\rm GeV}].
}}
\label{tab:massesA}
\end{table}

At {\underline {LC}} very precise mass values can be extracted 
from decay spectra and threshold scans \cite{r6A,r6B}.
The excitation curves for chargino production
in S-waves~\cite{r7} rise steeply with the velocity of the particles
near the thresholds and thus are very sensitive to their mass values;
the same is true for mixed-chiral selectron pairs in
$e^+e^-\to \tilde e_R^+ \tilde e_L^-$ 
and for diagonal pairs in 
$e^-e^-\to \tilde e_R^- \tilde e_R^-, \;  \tilde e_L^- \tilde e_L^-$
collisions, see Ref.\cite{r6B} which includes also the effects of
radiative corrections.  
Other scalar sfermions, as well as neutralinos, 
are produced generally in P-waves, with a
less steep threshold behaviour proportional to the
third power of the velocity.  Additional information,
in particular on the lightest neutralino $\tilde{\chi}^0_1$, can
be obtained from the very sharp edges of 2-body decay spectra,
such as ${\tilde{e}}^-_R \to e^- {\tilde{\chi}}^0_1$.

Typical mass parameters and the related measurement
errors are presented in Table~\ref{tab:massesA}:  
``LHC'' from LHC analyses and ``LC'' from LC analyses.
The third column ``LHC+LC'' presents the corresponding errors if the
experimental analyses are performed coherently, i.e. the
light particle spectrum, studied at LC with very high precision,
is used as input set for the LHC analysis.

\renewcommand{\arraystretch}{1.07}
\begin{table}[t]
\begin{center}
\begin{tabular}{|c||c|c|}
\hline
           & Parameter, ideal   & {``LHC+LC''} errors
\\ \hline\hline
 $M_1$        & 101.66   &   0.08  \\
 $M_2$        & 191.76   &   0.25  \\
 $M_3$        & 584.9    &   3.9   \\
\hline
$\mu$         & 357.4    &   1.3   \\
\hline  
 $M^2_{L_1}$  &$3.8191 \cdot 10^4$ & 82.   \\
 $M^2_{E_1}$  &$1.8441 \cdot 10^4$ & 15.   \\
 $M^2_{Q_1}$  &$29.67 \cdot 10^4$  & $0.32\cdot 10^4$ \\
 $M^2_{U_1}$  &$27.67 \cdot 10^4$ &  $0.86 \cdot 10^4$ \\
  $M^2_{D_1}$ &$27.45 \cdot 10^4$ &  $0.80 \cdot 10^4$ \\
$M^2_{H_2} $  &$-12.78 \cdot 10^4$& $0.11 \cdot 10^4$  \\
$A_t $        & $-497.$       &  9.    \\
\hline
$\tan\beta$   & 10.0              &  0.4   \\
\hline
\end{tabular}
\end{center}
\caption{{\it The extracted SUSY Lagrange mass and Higgs parameters 
at the electroweak scale in the reference point SPS1a
[mass unit {\rm GeV}].
}} 
\label{tab:params}
\end{table}

Mixing parameters must be extracted from measurements of cross
sections and polarization asymmetries, 
in particular from the production of chargino pairs and
neutralino pairs~\cite{r7}, both in diagonal or mixed form: 
$e^+e^- \rightarrow {\tilde{\chi}^+_i}{\tilde{\chi}^-_j}$
[$i$,$j$ = 1,2] and ${\tilde{\chi}^0_i} {\tilde{\chi}^0_j}$ 
[$i$,$j$ = 1,$\dots$,4]. 
The production cross sections for
charginos are binomials of $\cos\,2\phi_{L,R}$, the mixing angles
rotating current to mass eigenstates. Using polarized electron
and positron beams, the mixings can be determined in a model-independent
way.

The fundamental SUSY parameters can be derived to lowest order 
in analytic form: 
\begin{eqnarray}
\left|\mu\right|&=&M_W[\Sigma + \Delta[\cos2\phi_R+\cos2\phi_L]]^{1/2}
\nonumber\\
M_2&=&M_W[\Sigma - \Delta(\cos2\phi_R+\cos2\phi_L)]^{1/2}\nonumber\\
|M_1|&=& \left[ \textstyle \sum_i m^2_{\tilde{\chi}_i^0}  
                 -M^2_2-\mu^2-2M^2_Z\right]^{1/2}
\nonumber\\
|M_3|&=&m_{\tilde{g}} \nonumber\\
\tan\beta&=&\left[\frac{1+\Delta (\cos 2\phi_R-\cos 2\phi_L)}
           {1-\Delta (\cos 2\phi_R-\cos 2\phi_L)}\right]^{1/2} 
\label{eqn:basicLE}
\end{eqnarray}
where $\Delta = (m^2_{\tilde{\chi}^\pm_2}-m^2_{\tilde{\chi}^\pm_1})/(4M^2_W)$
and 
$\Sigma =  (m^2_{\tilde{\chi}^\pm_2}+m^2_{\tilde{\chi}^\pm_1})/(2M^2_W) -1$.
The signs of $\mu$, $M_{1,3}$ with respect to $M_2$ follow 
from similar relations
and from cross sections for ${\tilde{\chi}}$ production and $\tilde{g}$
processes. In practice one-loop corrections to the mass relations 
have been used to improve on the accuracy.  
 
The mass parameters of 
the sfermions are directly related to the
physical masses if mixing effects are negligible: 
\begin{equation}
m^2_{\tilde{f}_{L,R}}=M^2_{L,R}+m^2_f + D_{L,R} 
\end{equation}
with $D_{L} = (T_3 - e_f \sin^2 \theta_W) \cos 2 \beta \, m^2_Z$ 
and $D_{R} = e_f \sin^2 \theta_W \cos 2 \beta \, m^2_Z$ 
denoting the D-terms.  The non-trivial
mixing angles in the sfermion sector of the third generation
follow from the sfermion production cross sections  for 
longitudinally polarized e$^+$/e$^-$ beams, which are bilinear
in $\cos$/$\sin2\theta_{\tilde f}$.  The mixing angles and the two physical
sfermion masses are related to the tri-linear couplings $A_f$,
the higgsino mass parameter $\mu$ and $\tan\beta(\cot\beta)$
for down(up) type sfermions by:
\begin{equation}
A_f-\mu\tan\beta(\cot\beta)=
\frac{m^2_{\tilde{f}_1}-m^2_{\tilde{f}_2}}{2 m_f}\sin2\theta_{\tilde f}
\end{equation}
$A_f$ may be determined in the $\tilde{f}$ sector if $\mu$ has been 
measured in the chargino sector.

\renewcommand{\arraystretch}{1.07}
\begin{table}
\begin{center}
\begin{tabular}{|c||ccc|}
\hline \hspace*{1mm}
 & $\tilde q_R - \tilde\chi^0_1$
 & $\tilde l_L - \tilde\chi^0_1$ 
 & $m[\tilde g - [\tilde b_1]$                 \\\hline\hline
\SPHENO & 450.3 & 110.0 & 88.9 \\
$\Delta_{exp}^{LHC}$  
    & \phantom{0}10.9 
    & \phantom{00}1.6 
    & \phantom{0}1.8     \\ 
$\Delta_{th}\;\,$  
    & \phantom{00}8.1 
    & \phantom{000}0.23 
    & \phantom{0}6.8      \\\hline
\end{tabular}
\end{center}
\caption{{\it A sample of observable mass differences at LHC 
   for SPS1a and their 
   experimental ($\Delta_{exp}^{LHC}$) and present theoretical 
   ($\Delta_{th}$)  
   uncertainties due to variations of the SUSY scale. [All quantities in GeV]. 
   See also Ref.~\cite{Allanach:2003jw}}. 
\label{tab:LHCobs}}
\end{table}

Accuracies expected for the SUSY Lagrange parameters at the
electroweak scale for the reference point SPS1a are shown in
Table~\ref{tab:params}. They have been calculated by means of
SPheno2.2.0 \cite{Porod:2003um}. Theoretical errors, exemplified in
Table~\ref{tab:LHCobs}, have been estimated by varying the characteristic SUSY
scale between 100 GeV and 1 TeV. Note that these theoretical errors
do match the experimental LHC errors but they must be reduced by an
order of magnitude to match the expected accuracies at LC.

\section{Reconstruction of the Fundamental SUSY Theory}

The fundamental mSUGRA parameters [\ref{eq:sps1a}] at the GUT scale are
related to the low-energy parameters at the electroweak scale
by supersymmetric renormalization group 
transformations (RG)~\cite{RGE1,RGE2}
which to leading order generate the evolution for:\\
\begin{tabular}{ll}
 gauge couplings &: $\alpha_i = Z_i \, \alpha_U$ \hfill (5) \\
 gaugino masses  &: $M_i = Z_i \, M_{1/2}$ \hfill (6) \\
 scalar masses   &:   \\
 \multicolumn{2}{c}{$\hspace*{1cm} M^2_{\tilde\jmath} = M^2_0 + c_j M^2_{1/2} +
        \sum_{\beta=1}^2 c'_{j \beta} \Delta M^2_\beta$  \hfill (7)} \\
 trilinear  couplings &:  $A_k = d_k A_0   + d'_k M_{1/2}$         \hfill (8)
\end{tabular}
\refstepcounter{equation}
\refstepcounter{equation}
\label{eq:gaugino}
\refstepcounter{equation}
\label{eq:squark} 
\refstepcounter{equation}\\
The index $i$ runs over the gauge groups $i=SU(3)$, $SU(2)$, $U(1)$.
To leading order, the gauge couplings, and the gaugino and scalar mass
parameters of soft--supersymmetry breaking depend on the $Z$ transporters
\begin{eqnarray}
Z_i^{-1} =  1 + b_i \frac{\alpha_U}{ 4 \pi}
             \log\left(\frac{M_U}{ M_Z}\right)^2 
\end{eqnarray}
with $b[SU_3, SU_2, U_1] = -3, \, 1, \, 33 / 5$;
the scalar mass parameters depend 
also on the Yukawa couplings $h_t$, $h_b$, $h_\tau$
of the
top quark, bottom quark and $\tau$ lepton.
The coefficients $c_j$ 
for the slepton and squark doublets/singlets, 
and for the two Higgs doublets,
are linear combinations of the evolution
coefficients $Z$; the coefficients $c'_{j \beta}$ are of order unity. 
The shifts $\Delta M^2_\beta$, depending implicitly on all the other
parameters, are nearly zero for the first two families of 
sfermions but they can be rather large for the third family and for the 
Higgs mass parameters. 
The coefficients $d_k$ of the trilinear
couplings $A_k$ [$k=t,b,\tau$]  
depend on the corresponding Yukawa couplings 
and they are approximately unity for the
first two generations while being O($10^{-1}$) 
and smaller if the Yukawa couplings are
large; the coefficients $d'_k$, depending on gauge 
and Yukawa couplings, are of order unity.
Beyond the approximate solutions, the evolution equations 
have been solved numerically in the present analysis  to
two--loop order \cite{RGE2} and threshold effects have been
incorporated at the low scale \cite{bagger}.
The 2-loop effects as given in
Ref.~\cite{Degrassi:2001yf} have been included for
the neutral Higgs bosons and the $\mu$ parameter.

\begin{figure*}
\setlength{\unitlength}{1mm}
\begin{center}
\begin{picture}(160,57)
\put(-15.,-95){\mbox{\epsfig{figure=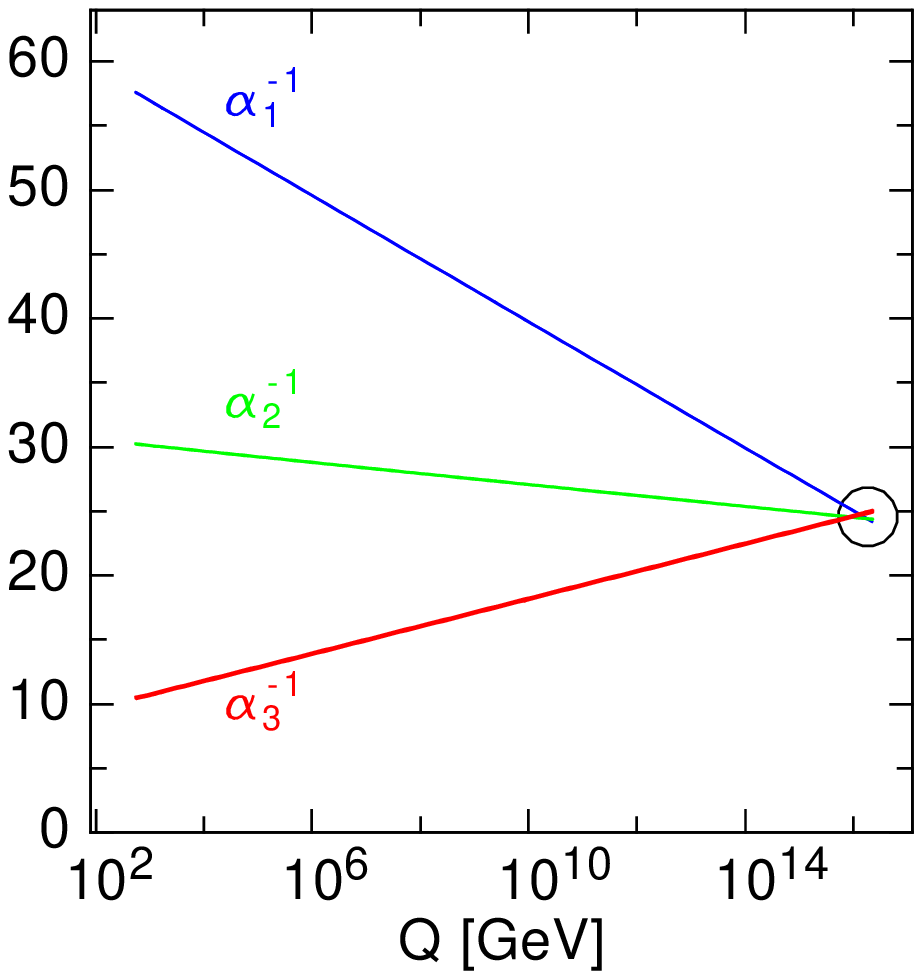,
                                   height=19.cm,width=14.cm}}}
\put(76,20){\mbox{\huge $\Rightarrow$}}
\put(65,-95){\mbox{\epsfig{figure=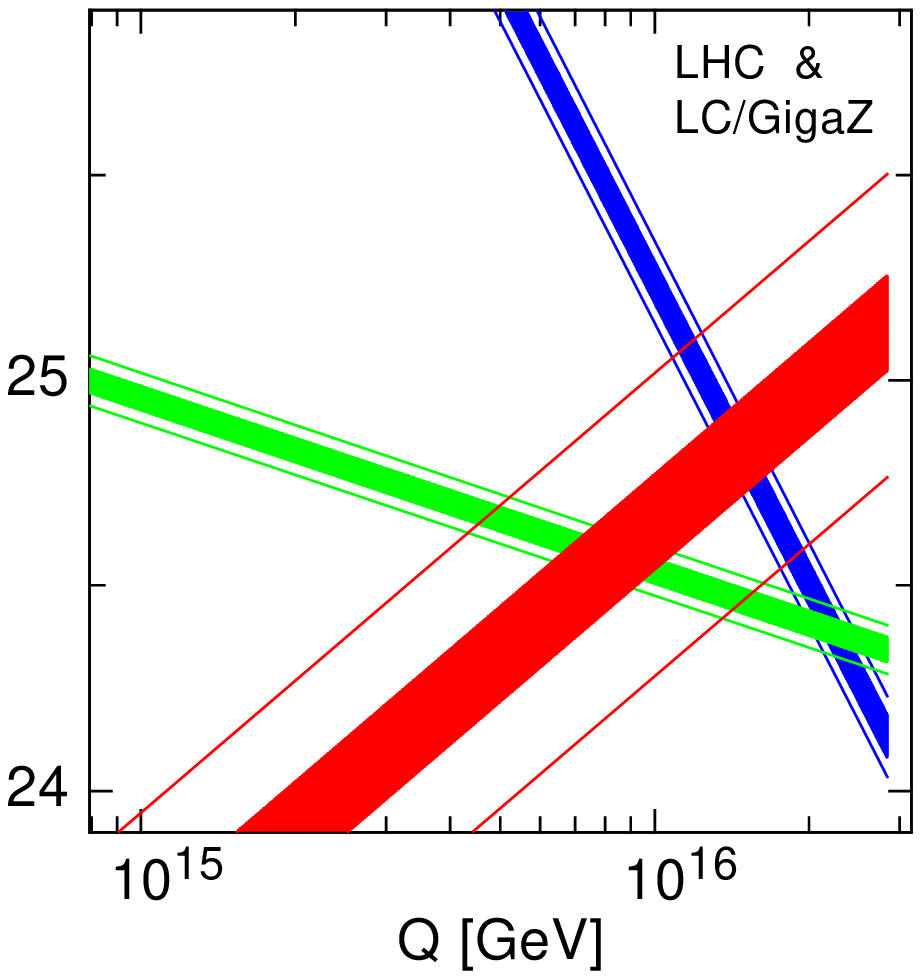,
                                   height=19.cm,width=14.cm}}}
\put(-3,59){\mbox{\bf a)}}
\put(84,59){\mbox{\bf b)}}
\end{picture}
\end{center}
\caption{{\it (a) Running of the inverse gauge couplings from low
  to high energies.
  (b) Expansion of the area around the unification point $M_U$ 
      defined by the meeting point of $\alpha_1$ with $\alpha_2$.
      The wide error bands are based on present data, and the spectrum
      of supersymmetric particles from LHC measurements within mSUGRA. 
      The narrow bands demonstrate the improvement expected by future
      GigaZ analyses \cite{Monig:2001hy} and the measurement of the complete
        spectrum at ``LHC+LC''.}}
\label{fig:gauge}
\end{figure*} 

\subsection{Gauge Coupling Unification}

Measurements of the gauge couplings at the electroweak scale
support very strongly the unification of the couplings at a scale
$M_U \simeq 2\times 10^{16}$~GeV \cite{r13A}.  
The precision, at the per--cent level, is
surprisingly high after extrapolations over
fourteen orders of magnitude in the energy 
from the electroweak scale to the grand unification scale $M_U$. 
Conversely, the
electroweak mixing angle has been predicted in this approach at the
per--mille level. The evolution of the gauge couplings from 
low energy  to the GUT scale $M_U$ has been carried out at two--loop accuracy
in the $\overline{DR}$ scheme.
The couplings are evolved to $M_U$ using 
2-loop RGEs \cite{RGE2}. 
The gauge couplings do not meet exactly, cf. Fig.~\ref{fig:gauge} and 
Tab.~\ref{tab:gauge}. 
The differences are to be attributed to high-threshold effects 
at the unification
scale $M_U$ and the quantitative evolution implies 
important constraints on the particle content at $M_U$
\cite{Ross:1992tz}.

\begin{figure*}
\setlength{\unitlength}{1mm}
\begin{center}
\begin{picture}(160,60)
\put(-2,-78){\mbox{\epsfig{figure=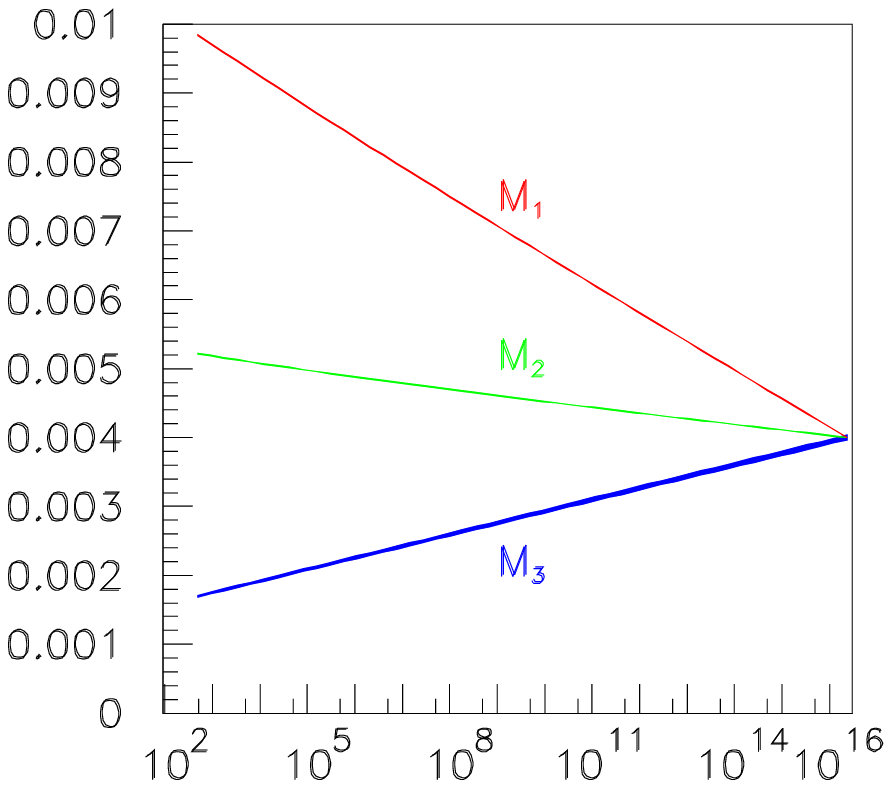,height=15cm,width=16cm}}}
\put(1,63){\mbox{\bf (a)}}
\put(12,61){\mbox{$1/M_i$~[GeV$^{-1}$]}}
\put(60,-5){\mbox{$Q$~[GeV]}}
\put(80,-78){\mbox{\epsfig{figure=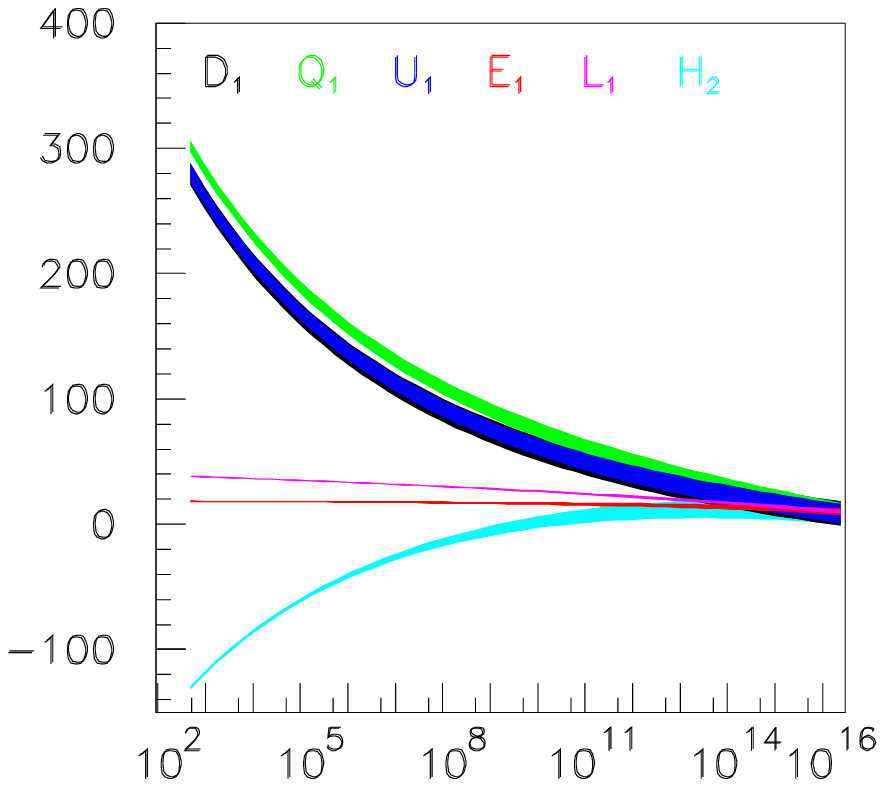,height=15cm,width=16cm}}}
\put(83,63){\mbox{\bf (b)}}
\put(94,61){\mbox{$M^2_{\tilde j}$~[$10^3$ GeV$^2$]}}
\put(142,-5){\mbox{$Q$~[GeV]}}
\end{picture}
\end{center}
\caption{{\it  Evolution, from low to high scales, (a) of 
the gaugino mass parameters
for ``LHC+LC'' analyses;
(b)  of the first/second generation sfermion mass parameters 
 and  
the Higgs mass parameter $M^2_{H_2}$.}   
}
\label{fig:sugra_LHC}
\end{figure*} 

\begin{table*}
\begin{center}
\begin{tabular}{|c||c|c|}
\hline
 & Present/''LHC'' & GigaZ/''LHC+LC'' \\
\hline \hline
$M_U$ & $(2.36 \pm 0.06)\cdot 10^{16} \, \rm {GeV}$ & 
           $ (2.360 \pm 0.016) \cdot 10^{16} \, \rm {GeV}$ \\
$\alpha_U^{-1}$ & $  24.19 \pm 0.10 $ &  $ 24.19 \pm 0.05 $\\ \hline
$\alpha_3^{-1} - \alpha_U^{-1}$ & $0.97 \pm 0.45$ & $0.95 \pm 0.12$ \\ \hline
\end{tabular}
\end{center}
\caption{{\it Expected errors on $M_U$ and $\alpha_U$ for the mSUGRA 
 reference point SPS1a, derived for the present level of 
 experimental accuracy and
 compared with expectations from GigaZ  \cite{Monig:2001hy}.
  Also shown is the difference between
 $\alpha_3^{-1}$ and $\alpha_U^{-1}$ at the unification point $M_U$.}}
\label{tab:gauge}
\end{table*}

\subsection{Gaugino and Scalar Mass Parameters}

In the bottom-up approach the fundamental supersymmetric theory
is reconstructed at the high scale from the available {\it corpus} of
experimental data without any theoretical prejudice. This approach exploits
the experimental information to the maximum extent possible and reflects an
undistorted picture of our understanding of the basic theory.
Such
a program can only be carried out in coherent ``LHC+LC'' analyses  while
the separate information from either machine proves insufficient.
The results for the evolution of the mass parameters from the electroweak
scale to the GUT
scale $M_U$ are shown in Fig.~\ref{fig:sugra_LHC}.  

On the left of Fig.~\ref{fig:sugra_LHC}
the evolution is presented for the 
gaugino parameters $M^{-1}_i$. It clearly
is under excellent control for the 
model-independent reconstruction
of the parameters and the test of universality
in the $SU(3) \times SU(2) \times U(1)$ group space.
In the same way the evolution of the scalar mass parameters can be
studied, presented in Figs.~\ref{fig:sugra_LHC}b 
for the first/second generation.
While the slepton parameters can be determined very accurately, 
the accuracy deteriorates for the squark parameters 
and the Higgs parameter $M^2_{H_2}$.

\section{Summary}

In supersymmetric theories stable
extrapolations can be performed from the electroweak scale
to the grand unification scale, close to the Planck scale.
This feature has been demonstrated compellingly in the evolution
of the three gauge couplings and of the soft supersymmetry breaking
parameters, which approach universal values at the GUT scale in minimal 
supergravity. 
The coherent ``LHC+LC'' analyses in which the measurements 
of SUSY particle
properties at LHC and LC mutually improve each other, result in a
comprehensive and detailed picture 
of the supersymmetric particle system. In particular, the gaugino sector
and the non-colored scalar sector are under excellent control.

This point can be highlighted by performing a global
mSUGRA fit of the universal parameters, c.f. Tab.~\ref{tab:univ_params}.
Accuracies at the level of per-cent to per-mille can be
reached, allowing us to reconstruct the structure of nature
at scales where gravity is linked with particle physics.

Though minimal supergravity has been chosen as a specific example, 
the method can
equally well be applied in other scenarios, such as 
left-right symmetric theories and 
superstring theories. The analyses offer the exciting opportunity to 
determine intermediate scales in left-right symmetric theories and to
measure effective string-theory parameters near the Planck scale.

\begin{table}
\begin{center}
\begin{tabular}{|c|c|c|}
\hline
                &  Parameter, ideal    & Experimental error \\ 
\hline\hline
$M_U$           & $2.36\cdot 10^{16}$ &  $2.2  \cdot 10^{14}$       \\
$\alpha_U^{-1}$ &   24.19          &     0.05     \\ \hline
$M_\frac{1}{2}$ & 250.             & 0.2     \\
$M_0$           & 100.             & 0.2     \\
$A_0$           & -100.            & 14      \\  
\hline
$\mu$           & 357.4            & 0.4     \\
\hline
$\tan\beta $    &  10.             & 0.4      \\  
\hline
\end{tabular}
\end{center}
\caption[]{\it Comparison of the ideal parameters with the
experimental expectations 
in the combined ``LHC+LC'' analyses
for the particular mSUGRA reference 
point adopted in this report [units in {\rm GeV}].} 
\label{tab:univ_params}
\end{table}


\end{document}